\documentclass[aoas,MSNbibl,nameyear,seceqn,dvips]{arximspdf}
\usepackage{algorithm}
\usepackage{dcolumn}
\usepackage{algorithmic}

\doi{10.1214/13-AOAS672} %kopijuoti is PTS
\volume{7}
\issue{4}
\pubyear{2013}
\firstpage{2293}
\lastpage{2314}

\makeatletter

\newcolumntype{d}[1]{D{.}{.}{#1}}

\newcommand{\var}{\operatorname{Var}}

\newcommand{\cal}{\mathcal}

\makeatother

\begin{document}
\begin{frontmatter}

\title{Heterogeneity and
behavioral response in continuous time capture--recapture, with
application to street cannabis use in Italy}
\runtitle{Continuous time recapture models and drug use in Italy}

\begin{aug}
\author[A]{\fnms{Alessio} \snm{Farcomeni}\corref{}\ead[label=e1]{alessio.farcomeni@uniroma1.it}}
\and
\author[B]{\fnms{Daria} \snm{Scacciatelli}\ead[label=e2]{daria.scacciatelli@gmail.com}}
\runauthor{A. Farcomeni and D. Scacciatelli}
\affiliation{Sapienza---University of Rome and University of Rome---Tor Vergata}
\address[A]{Department of Public Health\\
\quad and Infectious Diseases\\
Sapienza---University of Rome\\
Piazzale Aldo Moro, 5\\
00185 Rome\\
Italy\\
\printead{e1}}
\address[B]{Centre for Biostatistics and Bioinformatics\\
University of Rome---Tor Vergata\\
Via Orazio Raimondo 18\\
00173 Rome\\
Italy\\
\printead{e2}}
\end{aug}

% HISTORY:
\received{\smonth{11} \syear{2012}}
\revised{\smonth{7} \syear{2013}}

% ABSTRACT
%
\begin{abstract}
We propose a general and flexible capture--recapture model in continuous
time. Our model incorporates time-heterogeneity, observed and
unobserved individual heterogeneity, and behavioral response to
capture. Behavioral response can possibly have a delayed onset and a
finite-time memory. Estimation of the population size is based on the
conditional likelihood after use of the EM algorithm. We develop an
application to the estimation of the number of adult cannabinoid users
in Italy.
\end{abstract}

% KEYWORDS

\end{frontmatter}

%s1 #&#
\section{Introduction}

Capture--recapture experiments have been adopted in a wide range of
applications, including ecology, agriculture and veterinary science,
public health and medical studies, software engineering, behavioral
research and, in general, in the estimation of the size of hidden
populations.
%See for instance
Detailed reviews can be found in the International Working Group for
Disease Monitoring and Forecasting (\citeyear{iwg95a,iwg95b}) and
\citet{amstetal03}. Capture--recapture experiments are based on
repeatedly capturing subjects over time. The counting process of
captures is modeled so as to obtain an estimate of the number of
subjects never captured or, equivalently, of the size of the catchable
population. In discrete time there is a fixed number of capture
occasions. In continuous time each subject is at risk of capture in any
moment of a fixed-length period.
%For an account of applications of continuous time capture--recapture in
%ecology refer to \citet{wilsande95}.

In this paper we focus on continuous time models, motivated by a large
scale study on the population of drug users in Italy. The development
of new policies for tackling drug abuse is considered a concern in the
European Union.
%where action at EU level is essential.
There is extremely limited information on the number of cannabis users
in Italy. Population size estimates are usually based on indirect
estimation methods, for instance, through chemical analyses of waste
waters [\citet{EMCDDA2009}]. Our formal capture--recapture experiment
will provide a direct estimate at least for the catchable
subpopulation. Note that closed population capture--recapture models are
often used for estimation of the number of illicit drug users
[\citet
{Bohning2004,bouctrem05,chiang2007prevalence,vaissade2009capture,khazaei2012estimation}].
For a fixed time period, beginning with the introduction of a new law
on drug control, officers of different Italian police departments
identified and reported drug users. According to the new law, Art. 75,
possession of even a minimal quantity of any drug is a crime, which can
be punished with administrative sanctions. We have access to the entire
database, to the date of each capture, but only to part of the
information relating the drug user (e.g., sex, age, province), and no
information about the officer. The database records information on
different substances (cannabis, cocaine, heroin, other). In this paper
we focus on cannabis, whose prevalence of use in public places is the
largest and is in most cases the first drug used. See, for instance,
\citet{APD2009} and \citet{rossreyzuli11}. The study of cannabis use
is particularly important for planning prevention, evidence-based
interventions and understanding how abuses should be handled.
Furthermore, drug dealers often sell multiple drugs and the number of
drug dealers is expected to be proportional to the most requested drug
[\citet{reutklein86,bouctrem05}]. Therefore, a study on the
prevalence of cannabis may also be useful in planning actions for
tackling the illicit drug market.

Subjects were continuously at risk of being captured for the entire
observation period. As a consequence, we have a continuous time
capture--recapture experiment with observed covariates, possibly
unobserved heterogeneity, and time-heterogeneity. It is also reasonable
to allow for the possibility of arrests having consequences leading to
modifications of the future risk of capture, therefore having a
behavioural component in the model. Our catchable population is made of
cannabis users in Italy who use, buy and/or carry less than $500$ mg
of drug in the streets. It shall be noted that frequent and long-term
users mostly belong to this population, and these are at higher risk of
health consequences [e.g., \citet{semple2005cannabis}]. Drug use and/or
dealing in the streets causes a degree of public nuisance [e.g.,
\citet{jongwebe99}], generates stress and increases the risk of
psychological distress even to subjects that are not involved
[\citet{miroross03}]. We will also provide indirect estimates of the
entire population of cannabis users, that is, also including subjects
who use cannabis in private houses and never buy or carry it in the
streets.

With assumptions of time-homogeneity and absence of behavioral
response to capture, one could simply work with the individual
number of captures at the end of the observation period [e.g.,
\citet{Chao1987}]. %, Zelterman_1988, bohn:delr:08, rocc:et:al:11}.
When any of these two assumptions may not be met or shall be
verified, the conditional hazard function shall be explicitly
modeled through a Cox-type model [\citet{cox72}].
There are only few, but very neat, works dealing
with continuous time capture--recapture models along these lines.
%An approach based on sample coverage is proposed in
%These include \citet{Yip1996},\citet{Yip1999}, \citet{Chao2000}.
% propose a
%Cox-type model for inclusion of time-heterogeneity and observed
%covariates in the continuous time recapture context.
%an additive hazard model, also with time-heterogeneity and
%covariates.
%but allow for (long-term) behavioral
%responses, thus allowing the risk of being captured to change after
%the first capture.
The most relevant to our application are \citet{Chao2002},
who allow for time-heterogeneity, covariates and a behavioral
response; and \citet{Yip2007}, who also included a frailty in
the Cox-type model but no covariates.
%The latter is extended to
%covariates by \citet{stoketal11}.
%To our knowledge, the only work so far modeling
%unobserved heterogeneity in continuous time (together with
%time-heterogeneity and
%behavioral responses) is \citet{Yip2007}, who included a frailty in
%the Cox-type model.
In summary, the available approaches do not allow for
simultaneous modeling of two different kinds of individual heterogeneity:
the observed one, that can be modeled with covariates, and the
unobserved unmeasured heterogeneity, that shall be modeled through
a random frailty component.
Another limitation of the available models is
that the behavioral effect necessarily implies a \textit{long-term}
memory: the
hazard function is multiplied by a fixed factor after first capture,
and this factor does not vary over time. We believe that, especially
in our application in which a capture corresponds to a police officer
dealing with a drug user, individual response to capture could be more
complex. We generalize the long-term memory assumption
in two directions: first of all, we allow for a delayed onset of
the behavioral response; second, we allow this effect to
disappear after a fixed number of additional captures,
after a fixed time frame from the onset, or after the minimum
between the two conditions.
Not surprisingly, we will estimate a peculiar but easily
interpretable behavioral response on our data.

We denote our most general model by $M_{\mathrm{hotb}}$, where the $h$ stands
for the
unobserved heterogeneity, $o$ for the observed heterogeneity, $t$ for
the time-heterogeneity and $b$ for our flexible
behavioral response to capture. We use, as is common practice,
an unspecified baseline in the hazard function, thus nonparametrically modeling
time-heterogeneity. Unlike available models, we explicitly
distinguish and allow for the two different sources of individual
heterogeneity.
For estimation, we employ the conditional likelihood (and
Horvitz--Thompson estimator).
We set up a general expectation-maximization (EM) algorithm in order to
maximize the conditional likelihood, after employing an augmentation
scheme to tackle the issue of unknown frailty terms, and
drastically reducing the number of parameters involved by obtaining
an implicit estimate of the nonparametric baseline.
Notably, we also obtain a closed form expression for the integrals
involved, thus avoiding
numerical integration at the E-step.
Our model extends \citet{Yip2007} to the use of covariates. It can also
be seen as an
extension of the model of \citet{Chao2002} to inclusion of unobserved
heterogeneity. Finally, given our flexible
behavioral response to capture, we extend all previous works in this
direction in the spirit of the work of
\citet{farc11}, who proposes a general $M_b$
class of models in discrete time.
The rest of the paper is as follows: in the next section we
describe our $M_{\mathrm{hotb}}$ model and discuss all submodels in the class. In
Section~\ref{mb} we outline generalized behavioral responses.
In Section~\ref{inference} we show how to perform
inference on model parameters. We describe and analyze our data in
Section~\ref{data}, and
state conclusions in Section~\ref{concl}.

%s2 #&#
\section{General recapture models in continuous time}
\label{model}

Suppose we have a population of $N$ subjects, at risk of capture in
the time interval $[0,\tau]$. We assume the population is closed, that
is, there are no new users, no quitters and no migration during the
observation interval. We will discuss below the implications of these
assumptions for the data at hand. A common understanding is that if
$\tau$ is small enough, this assumption is safe.
Let $N_i(t)$, $i=1,\ldots,N$, denote the
number of times the $i$th subject has been captured in the time
interval $[0,t]$, $0\leq t \leq\tau$. We have $n$ subjects for which
$N_i(\tau)>0$, with captures at time $t_{ij}$, $i=1,\ldots,n$;
$j=1,\ldots,N_i(\tau)$.

Each $\{N_i(t); 0 \leq t \leq\tau\}$ is then a continuous time
counting process, along the lines of \citet{Chao2002} and \citet{Yip2007}.
We denote its intensity function, conditionally
on the capture history up to time $t$, by $\lambda_i(t)$.
Note that $\{N_i(t); 0 \leq t \leq\tau\}$ would be a Poisson process
under an assumption of absence of memory, that is, absence of
behavioral effects.
Let $Z_i$ denote a vector of subject specific covariates, of size $p$,
which is known only for the subjects captured at least once.
Our general $M_{\mathrm{hotb}}$ model can be stated as follows:
%
%e2.1 #&#
%
\begin{equation}
\label{hotb} \lambda_i(t) = \rho_i e^{\beta'Z_i}
\phi^{I(N_i(t)\geq1)} \omega(t),
\end{equation}
where $I(C)$ is the indicator function for condition $C$.
In (\ref{hotb}), $\beta$ denotes
a vector of log hazard-ratio coefficients describing
the observed heterogeneity;
$\omega(t)$~an unspecified nonnegative baseline function describing
the time-heterogeneity; $\phi$~measures a proportionality effect on the
risk after first capture, hence quantifying a behavioral response; and
$\rho_i$ is a subject-specific frailty term corresponding to
unobserved heterogeneity. All parameters, except for the vector
$\beta$, are assumed to be strictly positive. The frailty term is assumed
to arise from a distribution with support on ${\cal R}^+$. This
distribution shall not be left unspecified in general [see, e.g.,
\citet{link2003}], even if it is possible for certain models in
discrete time [Farcomeni and Tardella (\citeyear{farctard10,farctard09})].
A common choice [e.g., \citet{Yip2007}] is a
$\operatorname{Ga}(\alpha,\alpha)$ distribution, that is,
%
%e2.2 #&#
%
\begin{equation}
\label{fgamma} f(\rho_i) = \frac{\alpha^{\alpha}}{\Gamma(\alpha)} \rho
_i^{\alpha-1}
e^{-\rho_i\alpha}.
\end{equation}
This choice, that is, assuming that both parameters of the Gamma
distribution are identical, implies that a priori $E(\rho_i)=1$,
as with most frailty models, and $\var(\rho_i)=1/\alpha$.
As usual, a larger
variance corresponds to a more important role of unobserved
heterogeneity, that is,
a \textit{smaller} $\alpha$ can be interpreted as
a larger unobserved heterogeneity.
%We will compare this assumption with a more flexible
%one based on $\rho_i$ arising from a discrete distribution
%with a fixed number $k$ of support points.
%Hence we will also assume
%where $\xi_1=1$ for identifiability reasons (and obviously $\sum_j
%Both $\xi_j$, $j=2,\ldots,k$ and $\pi_j$, $j=1,\ldots,k$
%are unknown and shall be estimated using our
%conditional maximum likelihood approach, as we will outline below.
%The resulting model corresponds to a finite mixture of $M_{\mathrm{otb}}$
%models, equal
%up a multiplicative constant; and is particularly useful since it
%gives an easily interpreted clustering of the population with respect
%to unobserved heterogeneity. Such a discrete distribution is usually
%flexible enough to approximate
%even a complex true underlying distribution, but we can make no claim
%of adaptivity.
The model, as stated, involves the following restrictions: first,
we make the assumption of proportionality of hazards, that is, that
the effects of covariates included in the model are time-constant and
log-linear.
The formulation is furthermore restricted to time-constant covariates
as well. Second, we assume a multiplicative effect of the frailty
$\rho_i$, which
is also time constant; and we formulate a parametric assumption on its
distribution.
In (\ref{hotb}) we also make the usual long term memory assumption for the
behavioral effect. We will generalize (\ref{hotb}) to more complex
behavioral effects in Section~\ref{mb} below.
%
%t1 #&#
\begin{table}
\tablewidth=292pt
\caption{Models nested in $M_{\mathrm{hotb}}$ and the corresponding
restrictions needed to obtain them}
\label{mods}
\begin{tabular*}{\tablewidth}{@{\extracolsep{\fill}}lcc@{}}
\hline
\textbf{Model} & \textbf{Equation} & \textbf{Restrictions on} $\bolds{M_{\mathrm{hotb}}}$ \\
\hline
$M_{\mathrm{hotb}}$ & $\rho_i e^{\beta'Z_i} \phi^{I(N_i(t)\geq1)} \omega(t)$ &
\\
[1pt]
$M_{\mathrm{hob}}$ & $\rho_i e^{\beta'Z_i} \phi^{I(N_i(t)\geq1)} \omega$ &
$\omega(t) = \omega$ \\
[1pt]
$M_{\mathrm{htb}}$ & $\rho_i \phi^{I(N_i(t)\geq1)} \omega(t)$ & $\beta=0$ \\
[1pt]
$M_{\mathrm{hot}}$ & $\rho_i e^{\beta'Z_i} \omega(t)$ & $\phi=1$ \\
[1pt]
$M_{\mathrm{otb}}$ & $e^{\beta'Z_i} \phi^{I(N_i(t)\geq1)} \omega(t)$ &
$\rho_i=1$ \\
[1pt]
$M_{\mathrm{ho}}$ & $\rho_i e^{\beta'Z_i} \omega$ &
$\omega(t)=\omega$, $\phi=1$ \\
[1pt]
$M_{\mathrm{ht}}$ & $\rho_i \omega(t)$ &
$\beta=0$, $\phi=1$ \\
[1pt]
$M_{\mathrm{hb}}$ & $\rho_i \phi^{I(N_i(t)\geq1)} \omega$ &
$\beta=0$, $\omega(t)=\omega$ \\
[1pt]
$M_{\mathrm{ot}}$ & $e^{\beta'Z_i} \omega(t)$ &
$\phi=1$, $\rho_i=1$ \\
[1pt]
$M_{\mathrm{tb}}$ & $\phi^{I(N_i(t)\geq1)} \omega(t)$ &
$\beta=0$, $\rho_i=1$ \\
[1pt]
$M_{\mathrm{ob}}$ & $e^{\beta'Z_i} \phi^{I(N_i(t)\geq1)} \omega$ &
$\omega(t)=\omega$, $\rho_i=1$ \\
[1pt]
$M_h$ & $\rho_i \omega$ & $\omega(t)=\omega$, $\beta=0$, $\phi=1$ \\
[1pt]
$M_o$ & $e^{\beta'Z_i} \omega$ & $\omega(t)=\omega$, $\rho_i=1$,
$\phi=1$ \\
[1pt]
$M_t$ & $\omega(t)$ & $\beta=0$, $\rho_i=1$,
$\phi=1$\\
[1pt]
$M_b$ & $\omega\phi^{I(N_i(t)\geq1)}$ & $\omega(t)=\omega$, $\rho_i=1$,
$\beta=0$\\
[1pt]
$M_0$ & $\omega$ & $\omega(t)=\omega$, $\rho_i=1$,
$\beta=0$, $\phi=1$\\
\hline
\end{tabular*}
\end{table}
It is worth noticing here that straightforward assumptions can be used
to obtain
simpler classes of models, some of which are
commonly used in the capture--recapture
literature. Assuming, for instance, that $\omega(t)=\omega$ is constant
over time leads to time-homogeneity. Fixing $\beta=0$ leads to
assume that there is no observed heterogeneity, while assuming
$\rho_i=1$ corresponds
to no unobserved heterogeneity. Finally,
$\phi=1$ corresponds to no behavioral effects. An account of all
possible models and corresponding assumptions is given in Table~\ref{mods}.

%s3 #&#
\section{Flexible behavioral effects}
\label{mb}

Model (\ref{hotb}) includes
classical behavioral effects: the risk of capture changes,
and then remains constant after the first capture.
The assumption\vadjust{\goodbreak} that risk changes only at first capture may be
restrictive in some cases; see, for instance,
\citet{yangchao05,ramsseve10,farc11} and the references therein.
In continuous time capture--recapture, we outline two generalizations
of the usual $M_b$, which can be summarized in our modification
of (\ref{hotb}) as follows:
%
%e3.1 #&#
%
\begin{equation}
\label{hotbgen} \lambda_i(t) = \rho_i e^{\beta'Z_i}
\phi^{I (c_1 \leq N_i(t) \leq c_2\ \mathrm{and}\ t\leq t_{ic_1}+\Delta
_b )} \omega(t),
\end{equation}
where $c_1<c_2$ are fixed integers smaller than the maximum
number of recaptures $\max_i N_i(\tau)$, and $\Delta_b>0$.
The model is identifiable as soon as there is at least one
subject with at least $c_1+1$ captures, with the $c_1+1$th
in the interval $(t_{ic_1},t_{ic_1}+\Delta_b)$.
% for some
%$i=1,\ldots,n$, and that there is at least one subject with $c_1$
%captures.
Note that assuming $c_1=1$ and $c_2=\Delta_b=\infty$ gives back (\ref{hotb}).
In (\ref{hotbgen}) we allow for a delayed onset of the behavioral
response. For instance, $c_1=3$ means that a
behavioral effect is not expected after the first or second, but only
after the
third capture. Further, we allow for a finite time memory: after the
minimum between $c_2-c_1$ captures and a time frame of $\Delta_b$, the
risk of capture returns to the pre-behavioral-response state.
In our application, for instance, it is reasonable to expect that
$c_1=2$, given that more serious legal consequences are experienced
after the second capture, and that $c_2=\Delta_b=\infty$ (purely
delayed onset model). There would be a finite time behavioral
response if, for instance, a subject who is not captured
for a $\Delta_b$ time frame is forgiven the previous offenses.
A finite $c_2$ can be expected in applications of
capture--recapture to animal populations, in which the animals may get
used to traps after being trapped a few times.

%s4 #&#
\section{Inference}
\label{inference}

Given that we cannot measure covariates for subjects never captured,
the only practical possibility
for inference is via maximization of the conditional
likelihood, that is, the likelihood obtained conditioning on the event
that $N_i(\tau)>0$. The estimated parameters are then used to build an
Horvitz--Thompson (HT) type estimator for $N$
[\citet{horvthom52,Sanaesti1972}].
For the sake of conciseness and simplicity of notation
we here outline inference for model $M_{\mathrm{hotb}}$ in
(\ref{hotb}). The strategy for submodels can be found along similar
lines.

In what follows, let $\Omega(t) = \int_0^t \omega(s)\,ds$
and
$\gamma_i = e^{\beta'Z_i}$. Denote the
probability of having at least one capture for individual $i$,
conditional on $\rho_i$, as
%
%e4.1 #&#
%
\begin{equation}
\label{pi} P_i=1-e^{-\rho_{i}\gamma_{i}\Omega(\tau)}.
\end{equation}
With an argument similar to the reasoning in
\citet{crowetal91} and \citet{Chao2002},
we can thus obtain a likelihood contribution for the $i$th subject,
conditional on $\rho_i$, as follows:
\[
L_i\propto\phi^{N_i(\tau)-1}\gamma_i^{N_i(\tau)} \Biggl[\prod_{j=1}^{N_i(\tau)}
\omega(t_{ij}) \Biggr] \rho_{i}^{N_i(\tau)}e^{-\rho_i\gamma_i \Omega^*_i(\tau
)} /P_i,
\]
where
%
%e4.2 #&#
%
\begin{equation}
\label{omegastar} \Omega^*_i(\tau) = \phi\Omega(\tau)+(1-\phi)
\Omega(t_{i1}).\vadjust{\goodbreak}
\end{equation}

When $\rho_i$ is a random effect arising from a parametric
distribution $F(\rho_i)$, we shall integrate the above expressions with
respect to this distribution. We consequently have a likelihood that
is conditional on having at least one capture, and marginal
with respect to random effects.
The expression for this likelihood is
%
%e4.3 #&#
%
\begin{equation}
\label{condlik}\quad  L_c = \prod_{i=1}^n
\phi^{N_i(\tau)-1}\gamma_i^{N_i(\tau)} \Biggl[\prod
_{j=1}^{N_i(\tau)} \omega(t_{ij}) \Biggr] \int
_0^{\infty} \frac{x^{N_i(\tau)}
e^{-x\gamma_i \Omega^*_i(\tau)}}{1-e^{-x\gamma_i\Omega(\tau)}}\,dF(x).
\end{equation}
The above expression is a function of the covariate parameters
$\beta$, the behavioral parameter $\phi$,
the baseline function $\omega(t)$
and any parameter involved in $F(\rho_i)$.
Estimates of the above parameters can be obtained by maximizing
$L_c$. Note that this maximum likelihood estimator will be consistent
and asymptotically normal given that it satisfies properties in
\citet{nielgillandesore92} and \citet{Sanaesti1972}. See also
\citet{gill92}.

%When $F(\rho_i)$ is a discrete random variable, the integral reduces to
%a sum of $k$ terms.
When $F(\rho_i)$ is a $\operatorname{Ga}(\alpha,\alpha)$,
we can solve the integral and obtain a closed form expression for
$L_c$. Note, in fact, that under the
assumption of Gamma distributed random effects
\[
L_c = \prod_{i=1}^n
\phi^{N_i(\tau)-1}\gamma_i^{N_i(\tau)} \Biggl[\prod_{j=1}^{N_i(\tau)}
\omega(t_{ij}) \Biggr] \frac{\alpha^{\alpha}}{\Gamma(\alpha)}
\int_0^{\infty} \frac{x^{N_i(\tau)+\alpha-1}
e^{-x(\gamma_i \Omega^*_i(\tau)+\alpha)}}{1-e^{-x\gamma_i\Omega(\tau
)}}\,dx.
\]
The integral is proportional to the integral
definition of the generalized Zeta function [\citet{gradryzh07}, Section~3.411,
formula 7] $\zeta(s,a)$, which is defined
as the series $\sum_n 1/(n+a)^s$, and can be
evaluated simply. See \citet{magnobersoni66} for details.

It can be shown that
%
%e4.4 #&#
%
\begin{equation}
\label{zetadef} \int_0^{\infty} x^a
e^{-bx}/\bigl(1-e^{-cx}\bigr)\,dx = \frac{\Gamma(a+1)}{c^{a+1}}
\zeta(a+1,b/c).
\end{equation}
Hence, after straightforward algebra we obtain that, under assumption
of Gamma distributed random effects,
%
%e4.5 #&#
%
\begin{equation}
\label{closedform} L_c = \prod_{i=1}^n
L_i,
\end{equation}
where
%
%e4.6 #&#
%
\begin{eqnarray}
\label{li} L_i &=& \phi^{N_i(\tau)-1}\gamma_i^{N_i(\tau)}
\Biggl[\prod_{j=1}^{N_i(\tau)}
\omega(t_{ij}) \Biggr] \frac{\alpha^{\alpha}\Gamma(\alpha+N_i(\tau
))}{\Gamma(\alpha)(\gamma_i
\Omega(\tau))^{\alpha+N_i(\tau)}} \nonumber\\[-8pt]\\[-8pt]
&&{}\times\zeta\biggl(
\alpha+N_i(\tau),\frac{\gamma_i
\Omega^*_i(\tau)+\alpha}{\gamma_i\Omega(\tau)} \biggr).\nonumber
\end{eqnarray}

%s4.1 #&#
\subsection{Maximization of the conditional likelihood}

Even if, with Gamma distributed random effects,
we have a closed-form expression for the conditional
likelihood, its direct maximization is cumbersome,\vadjust{\goodbreak}
as it would require
a precise evaluation of the derivatives of the generalized Zeta
function, which are not available in closed form.
More importantly, our approach would be limited to the assumption of
Gamma distributed random effects.
We give in this section a maximization strategy
based on the EM algorithm, which can
be easily adapted to any distributional assumption on $\rho_i$.

The EM algorithm proceeds by iterating two steps. At the E-step we
compute the conditional expected values of the frailty terms and plug
them into the \textit{complete} conditional likelihood, that is, the
likelihood we would have if we could observe the frailty terms.
This likelihood can be expressed as\looseness=-1
%
%e4.7 #&#
%
\begin{equation}
\label{complik} {\tilde L}_c = \prod_{i=1}^n
{\tilde L}_i,
\end{equation}\looseness=0
where
%
%e4.8 #&#
%
\begin{eqnarray}
\label{litilde} {\tilde L}_i&=&\phi^{N_i(\tau)-1}
\gamma_i^{N_i(\tau)} \Biggl[\prod_{j=1}^{N_i(\tau)}
\omega(t_{ij}) \Biggr] \rho_i^{N_i(\tau)}
e^{-\rho_i\gamma_i \Omega^*_i(\tau)}/\bigl(1-e^{-\rho_i\gamma_i\Omega
(\tau)}\bigr) f(\rho_i),\hspace*{-31pt}
\end{eqnarray}
where $f(\cdot)$ denotes the density or pmf of the random effects.

More precisely, at the E-step we substitute $\rho_i$ in
(\ref{complik}) with
%
%e4.9 #&#
%
\begin{equation}
\label{hatrho} \hat\rho_i = \int_0^{\infty}
x \,dF\bigl(x|N_i(t),Z_i, \beta,\omega(t),\phi\bigr).
\end{equation}
Note that (\ref{litilde}) is not linear in $\rho_i$, hence, its
conditional expectation does not coincide with its value at
(\ref{hatrho}). The E-step we propose is therefore only
approximate. The approximation is usually good, as we work with the
logarithm of (\ref{complik}), which is approximately linear for
large intervals, and due to the fact that
$\rho_i$ does not change much from one iteration to another.
%In our applications the
%likelihood was always nondecreasing at the plug-in E-step.
If a decreasing likelihood is observed, the plug-in E-step shall be
substituted with a Monte Carlo E-step.

The conditional density for the frailty terms can be obtained through
the Bayes theorem as follows:
%
%e4.10 #&#
%
\begin{eqnarray}
\label{posterior}
&&f\bigl(\rho|N_i(t),Z_i,\beta,
\omega(t),\phi\bigr) \nonumber\\
&&\qquad= \frac{f(N_i(t)|\rho,Z_i,\beta,\omega(t)\phi
)f(\rho)}{\int_0^{+\infty}
f(N_i(t)|\rho,Z_i,\beta,\omega(t)\phi)f(\rho)\,d\rho} \\
&&\qquad= \frac{{\tilde
L}_i}{\int_0^{\infty} {\tilde L}_i\,d\rho}.\nonumber
\end{eqnarray}
%
%When the frailty arises from a discrete distribution,
%$\hat\rho_i$ can be simply computed as the ratio of two $k$-term
%sums, since $f(\rho)=\pi_j$ when $\rho=\xi_j$; $j=1,\ldots,k$ and
%$f(\rho)=0$ when $\rho\neq\xi_j$ for any $j=1,\ldots,k$.
When the frailty is assumed to arise from a Gamma distribution, we have
a closed-form expression for $\hat\rho_i$.
The resulting $\hat\rho_i$ is a ratio, whose
denominator corresponds
to $L_i$ as defined in (\ref{li}) and, with similar
calculations\vadjust{\goodbreak}
involving (\ref{zetadef}), whose nominator corresponds to
\begin{eqnarray*}
&&\phi^{N_i(\tau)-1}\gamma_i^{N_i(\tau)} \Biggl[\prod_{j=1}^{N_i(\tau)}
\omega(t_{ij}) \Biggr] \frac{
\alpha^{\alpha} \Gamma(\alpha+N_i(\tau)+1)}{\Gamma(\alpha) (\gamma_i
\Omega(\tau))^{\alpha+N_i(\tau)+1}}\\
&&\qquad{}\times
\zeta\biggl(\alpha+N_i(\tau)+1,\frac{\gamma_i
\Omega^*_i(\tau)+\alpha}{\gamma_i\Omega(\tau)} \biggr).%
\end{eqnarray*}
After simplification of the terms, and noting that
$\Gamma(x+1)=x\Gamma(x)$, we get
%
%e4.11 #&#
%
\begin{equation}
\label{rhoi} \hat\rho_i = \frac{\alpha+N_i(\tau)}{\gamma_i\Omega(\tau
)} \frac{\zeta(\alpha+N_i(\tau)+1,({\gamma_i
\Omega^*_i(\tau)+\alpha})/({\gamma_i\Omega(\tau)}) )}{\zeta(\alpha+N_i(\tau
),({\gamma_i
\Omega^*_i(\tau)+\alpha})/({\gamma_i\Omega(\tau)}) )}.
\end{equation}
Other distributional assumptions may lead to the need of numerical
integration methods at the E-step or to the use of MCMC in order to
sample from the posterior distribution of the random effects [therefore
obtaining an MCEM algorithm, see, e.g., \citet{booth99}]. In the
latter case, the $M$-step which we will describe now can be still used,
with minor adjustments. After computation of (\ref{hatrho}), we plug it
in (\ref{complik}) to obtain an Expected Complete Conditional
Likelihood (ECCL). The $M$-step consists in maximizing the ECCL with
respect to $\beta$, $\omega(t)$, $\phi$ and any parameter involved in
the random effect distribution $F(\rho_i)$. This step is particularly
cumbersome due to the fact that there is an extremely large number of
parameters involved in the ECCL when the baseline hazard function
$\omega(t)$ is not known and not assumed to be constant. To tackle this
issue, we note that the nonparametric MLE for $\Omega(t)$ is a step
function, with jumps occurring at capture times, that is, the estimated
baseline hazard can be expressed as
\[
\Omega(t) = \sum_k \theta_k
I(t_{(k)} \leq t ),
\]
where $\theta_k=\omega(t_{(k)})$ measures the
size of the jump at the $k$th capture time. Maximizing the conditional
likelihood with respect to $\theta_k$, we obtain a Nelson--Aalen type
estimator [\citet{nels72,aale78}]. See also \citet{Chao2002}
for a similar reasoning in the continuous capture--recapture context. A
closed-form expression for $\hat\theta_k$ can be obtained by computing
the first derivative of the log-ECCL with respect to $\theta_k$ and
equating to zero. Straightforward algebra gives
\[
0 = \frac{dN(t_{(k)})}{\theta_k}-\sum_{i=1}^n{\hat\rho}_i \gamma_i \biggl[\phi
+(1-\phi) I(t_{(k)}<t_{i1})+\frac{e^{-{\hat\rho}_i\gamma_i\Omega(\tau
)}}{1-e^{-{\hat\rho}_i\gamma_i\Omega(\tau)}} \biggr],
\]
where $dN(t_{(k)})$ gives the number of captures occurring exactly at
the $k$th capture time. Consequently,
%
%e4.12 #&#
%
\begin{equation}
\label{hattk} \hat\theta_k = \frac{dN(t_{(k)})}{\sum_{i=1}^n \hat\rho
_i \gamma_i [
\phi^{I(t_{i1} < t_{(k)})}+{e^{-{\hat\rho}_i\gamma_i\Omega(\tau
)}}/({1-e^{-{\hat\rho}_i\gamma_i\Omega(\tau)}}) ]}.
\end{equation}
The resulting baseline hazard estimator can be seen as a Nelson--Aalen
type estimator with covariates, with an exponential term correction
at the denominator obtained as a consequence of conditioning to subjects
with at least one event.

The expression for (\ref{hattk}) shall be now
plugged in the ECCL, thereby drastically reducing the number of
parameters to $\Omega(\tau)$, $\phi$, $\beta$ and any parameter
involved in the random effect distribution $F(\rho_i)$.
These are estimated by solving a system of equations
%In order to estimate $\Omega(\tau)$, we note that
%which gives a nonlinear equation for $\hat\Omega(\tau)$.
which are obtained by equating to zero
the first derivatives of the ECCL after plug-in of $\hat\theta_k$,
plus an additional equation due to the constraint
%
%e4.13 #&#
%
\begin{equation}
\label{constrOmega} \sum_k \hat
\theta_k = \hat\Omega(\tau).
\end{equation}
Let now, for ease of notation,
%
%e4.14 #&#
%
\begin{equation}
\label{A} A\bigl(\gamma_{h},{\hat\rho_{h}},\Omega(\tau)
\bigr)=\frac{ e^{-{\hat\rho}_h\gamma_h\Omega(\tau)}}{1-e^{-{\hat\rho
}_h\gamma_h\Omega(\tau)}}
\end{equation}
and
%
%e4.15 #&#
%
\begin{equation}
\label{B} B\bigl(\phi,\gamma,\hat\rho,\Omega(\tau),t_k\bigr)= \sum
_{h=1}^n{\hat\rho}_h
\gamma_h \biggl[\phi^{I(t_k>t_{h1})}+\frac{e^{-{\hat\rho}_h\gamma
_h\Omega(\tau)}}{1-e^{-{\hat\rho}_h\gamma_h\Omega(\tau)}} \biggr].
\end{equation}

For what concerns the derivative with respect to $\phi$, we obtain
%
%e4.16 #&#
%
\begin{eqnarray}
\label{eqphi}
&&\sum_{i=1}^n\Biggl\{
\frac{(N_i(\tau)-1)}{\phi}\nonumber\\
&&\qquad{}-\sum_{j=1}^{N_i(\tau)}
\frac{\sum_{p=1}^n {\hat\rho}_p\gamma_p
I(t_{ij}>t_{p1})}{B(\phi,\gamma,{\hat\rho},\Omega(\tau),t_{ij})}
\nonumber\\[-8pt]\\[-8pt]
&&\qquad\hspace*{0pt}{}-{\hat\rho}_i\gamma_i\sum
_k\frac{d N(t_k)}{
B(\phi,\gamma,{\hat\rho},\Omega(\tau),t_{k})} \nonumber\\
&&\hspace*{70pt}{}\times\biggl[{I(t_k>t_{i1})}-
\phi^{I(t_k>t_{i1})}\frac{\sum_{p=1}^n{\hat\rho}_p\gamma_p{I(t_k>t_{p1})}}{
B(\phi,\gamma,{\hat\rho},\Omega(\tau),t_{ij})}
\biggr]\Biggr\};\nonumber
\end{eqnarray}
while for what concerns $\Omega(\tau)$, we have
%
%e4.17 #&#
%
\begin{eqnarray}
\label{eqO} &&\sum_{i=1}^n\Biggl\{\sum _{j=1}^{N_i(\tau)}
\biggl[\frac{\sum_{k=1}^n
e^{{\hat\rho}_k\gamma_k\Omega(\tau)}({\hat\rho}_k\gamma_k
A({\gamma_{k},{\hat\rho_{k}},\Omega(\tau)}))^2}{B(\phi,\gamma,\hat\rho,\Omega(\tau),t_{ij})}
\biggr]
\nonumber\\
&&\qquad\hspace*{0pt}{}-{\hat\rho}_i\gamma_i \Biggl[\sum
_k\phi^{I(t_k>t_{i1})} \frac{dN(t_k)}{B(\phi,\gamma,{\hat\rho},\Omega
(\tau),t_{k})^2} \nonumber\\[-8pt]\\[-8pt]
&&\qquad\hspace*{35.7pt}{}\times\sum
_{p=1}^n e^{{\hat\rho}_p\gamma_p\Omega(\tau)}\bigl({\hat
\rho}_p\gamma_p A\bigl(\gamma_{p},{\hat
\rho_{p}},\Omega(\tau)\bigr)\bigr)^2 \Biggr]
\nonumber\\
&&\hspace*{145pt}{}+{\hat\rho}_i\gamma_i A\bigl(
\gamma_{i},{\hat\rho_{i}},\Omega(\tau)\bigr)\Biggr\}.\nonumber
\end{eqnarray}
Finally, taking the first derivative of the log-ECCL with respect to
$\beta_h$, $h=1,\ldots,p$, we obtain the $p$ equations
%
%e4.18 #&#
%
\begin{eqnarray}
\label{eqbeta}
\hspace*{-4pt}&&\sum_{p=1}^n \Biggl\{
\frac{N_p(\tau)}{\gamma_p}
-\sum_{i=1}^n\sum
_{j=1}^{N_i(\tau)}\bigl(\bigl[{\hat\rho}_p
A\bigl({\gamma_{p}},{\hat\rho_{p}},\Omega(\tau)\bigr)\nonumber\\
\hspace*{-4pt}&&\hspace*{108pt}{}\times\bigl(1-{\hat\rho}_p\gamma_p
\Omega(\tau) {e^{{\hat\rho}_p\gamma_p\Omega(\tau)}}
A\bigl({\gamma_{p}},{\hat\rho_{p}},\Omega(\tau)\bigr)\bigr)\nonumber\\
\hspace*{-4pt}&&\hspace*{244.5pt}{}+{\hat\rho}_p\phi
^{I(t_{ij}>t_{p1})}\bigr]\nonumber\\
\hspace*{-4pt}&&\hspace*{104.5pt}{}/B\bigl(\phi,\gamma,{{\hat\rho}},\Omega(\tau),t_{ij}\bigr)\bigr)
\\
\hspace*{-4pt}&&\quad\qquad\hspace*{0pt}{}-{\hat\rho}_p \biggl[\sum_k
\phi^{I(t_k>t_{p1})}\frac{dN(t_k)}{{B({\gamma,{{\hat\rho}},\Omega(\tau
),t_k})}} \biggr]\nonumber
\\
\hspace*{-4pt}&&\quad\qquad\hspace*{0pt}{}+\sum_{i=1}^n {\hat
\rho}_i \gamma_i \biggl\{\sum
_k\phi^{I(t_k>t_{i1})}\frac{dN(t_k) {\hat\rho}_p }{({B({\phi,\gamma,{{\hat\rho}},\Omega(\tau),t_k})})^2}\nonumber\\
\hspace*{-4pt}&&\quad\hspace*{88.1pt}{}\times \bigl[
\phi^{I(t_k>t_{p1})}+ A\bigl({\gamma_{p}},{\hat\rho_{p}},
\Omega(\tau)\bigr)\nonumber\\
\hspace*{-4pt}&&\quad\hspace*{103pt}{}-{\hat\rho}_p\gamma_p\Omega(\tau)
e^{{\hat\rho}_p\gamma_p\Omega(\tau)} \bigl(A\bigl({\gamma_{p}},{\hat\rho_{p}},
\Omega(\tau)\bigr)\bigr)^{2}\bigr]\biggr\}\nonumber\\
\hspace*{-4pt}&&\hspace*{195.6pt}{}+{\hat\rho}_p
\Omega(\tau) A\bigl({\gamma_{p}},{\hat\rho_{p}},\Omega(\tau)
\bigr)\Biggr\} \gamma_p Z_{ph},\nonumber
\end{eqnarray}
where $Z_{ph}$ denotes the $ph$th entry of the covariate matrix $Z$.
The algebra involved in obtaining (\ref{eqphi}), (\ref{eqO}) and
(\ref{eqbeta}) is given in the supplement [\citet{farcscac13}].
Note that combining (\ref{eqphi}) with (\ref{eqO}) and
(\ref{eqbeta}) we obtain the score vector related to $\phi$,
$\Omega(\tau)$ and $\beta$ in closed form.
In order to conveniently proceed with the $M$-step,
we exploit the Newton--Raphson (NR) algorithm.
% with starting
%solution given by the current parameter values.
The NR algorithm only involves numerically
computing the first derivative of the score vector above, augmented
with (\ref{constrOmega}).
Parameters involved in the random effects distribution can be
tackled separately.
%When a ${\rm Ga}(\alpha,\alpha)$ distribution is
%assumed, %one can simply obtain $\hat\alpha$ using the method of
%moments as the
%inverse of the variance of $\hat\rho_i$, $i=1,\ldots,n$.
%Alternatively,
%it is straightforward to check that the score for
%$\alpha$ can be expressed as
%$n\log(\alpha)+n-n\psi(\alpha)+\sum_{i=1}^n \log(\hat\rho_i) -\hat
%where $\psi(\cdot)$ denotes the first derivative of the logarithm of
%the Gamma function,
%and the derivative of the score is
%$n/\alpha-n\psi'(\alpha).$
We can summarize our algorithm at the $t$th iteration
with the following pseudo-code, which is iterated until convergence
in the likelihood.

\begin{algorithm}
\begin{algorithmic}
\STATE Update $\hat\rho_i^{(t)}$ as in (\ref{rhoi}).
\STATE Solve the system of equations given by (\ref{constrOmega}),
(\ref{eqphi}), (\ref{eqO}) and (\ref{eqbeta}) to update $\hat\Omega(\tau)$,
$\hat\beta$ and $\hat\phi$.
\STATE Update $\hat\alpha$ as the maximizer
of $\prod_i f(\hat\rho_i)$, where $f(\rho_i)$ is defined in
(\ref{fgamma}).
\STATE Compute $\hat\theta^k$ as in (\ref{hattk}).
\STATE Compute the likelihood as in (\ref{closedform}).
\end{algorithmic}
\end{algorithm}

Final comments concern computation of the information matrix,
model choice and hypothesis testing.
Regarding the first issue, we note that
minus the derivative of the score vector
corresponds to the observed information matrix.
Since the score vector is available
in closed form, we can obtain the corresponding information
matrix as a natural by-product of our maximization strategy.
The observed information matrix at the maximum likelihood estimate
can be used to compute the standard errors in the usual way.
A~similar approach for estimation of the observed information matrix
after use of the EM algorithm is proposed in \citet{bartfarc09}, in a
completely
different context. A simulation study in \citet{bartfarc09} gives
evidence of the validity
of this procedure. The standard errors can then be used to build Wald
statistics for testing.
%The value of the likelihood at the MLE can be used for model
%choice, using for instance the Akaike Information Criterion (AIC),
%see \citet{akai73} and \citet{burnwhitande1995}.
%An issue with the above expression is that it is a function of $\hat
%captured. This issue is easily overcome noting that these appear only
%in the first sum, which is constant over $k$. We can consequently
%solve this issue simply by introducing a new parameter $\eta$,
%and letting
% (\phi-1) \sum_{i: t_{i1} < t_{(k)}} \hat\rho_i \gamma_i}.
%We have thus drastically reduced the number of parameters, since after
%pluggin (\ref{hattheta}) into (\ref{complik}), we need only maximize
%it with respect to $\phi$, $\beta$ and $\eta$.

%s4.2 #&#
\subsection{Other assumptions on the behavioral effect}

We briefly sketch in this paragraph how to modify the EM algorithm
proposed for model $M_{\mathrm{hotb}}$ when more flexible assumptions are used
for the behavioral effects, as summarized in equation
(\ref{hotbgen}).
In order to estimate such a model, it suffices to substitute the
indicator function arising from (\ref{hotb})
with the more complex indicator function involved in (\ref{hotbgen}).
These substitutions occur directly
in (\ref{hattk}), (\ref{B}),
(\ref{eqphi}), (\ref{eqO}) and (\ref{eqbeta}).
Further, straightforward algebra leads to redefine $\Omega^*_i(\tau)$ as
%
%e4.19 #&#
%
\begin{equation}
\label{omegastar2} \Omega^*_i(\tau) = \Omega(\tau)+(1-\phi) \bigl(
\Omega(t_{ic_1})-\Omega\bigl(\min(t_{ic_2},t_{ic_1}+
\Delta_b)\bigr)\bigr).
\end{equation}
Also, the likelihoods are modified:
in (\ref{condlik}), (\ref{li}) and (\ref{litilde}),
$\phi^{N_i(\tau)-1}$ shall be substituted with
\[
\phi^{\sum_{j=c_1}^{c_2} I(t_{ij}<=t_{ic_1}+\Delta_b)},
\]
where it is understood that $t_{ij}=\infty$ if $j>N_i(\tau)$.

Recall that $c_1,c_2$ and $\Delta_b$ must be
fixed a priori.
In order to carefully choose these parameters,
we can fit models for a range of values of $c_1$,
$c_2$ and $\Delta_b$ and
use the one corresponding to the largest log-likelihood.
The selected combination $(c_1,c_2,\Delta_b)$
is important information, telling
us, for instance, that a behavioral response
is not expected before the $c_1$th capture.

%s4.3 #&#
\subsection{Estimation of population size}

The maximum likelihood estimates can be used in a final E-step, thus
obtaining the corresponding $\hat\rho_i$. The latter can be used to
maximize the residual likelihood [see \citet{Sanaesti1972,farctard09}]
through a Hortvitz--Thompson estimator of the kind
%
%e4.20 #&#
%
\begin{equation}
\label{ht} \hat N = \sum_{i=1}^n
\frac{1}{1-e^{-\hat\rho_{i}\hat\gamma_{i}\hat\Omega(\tau)}}.
\end{equation}
When $N$ is large, as in our application, parameters involved in
(\ref{ht}) are approximately multivariate normal, and their covariance
matrix can be estimated as described above.

In order to obtain the standard error of $\hat N$, we note
that $n$ is random as well and that (\ref{ht}) should be
actually expressed as
\[
\hat N = \sum_{i=1}^N \frac{\delta_i}{1-e^{-\hat\rho_{i}\hat\gamma
_{i}\hat\Omega(\tau)}},
\]
where the random variable $\delta_i$ is defined
as $\delta_i=I(N_i(\tau)\geq1)$, $i=1,\ldots,N$.
Therefore [\citet{vandetal03,bohn08}], by conditioning,
%
%e4.21 #&#
%
\begin{equation}
\label{condV} \var(\hat N) = \var_n\bigl(E(\hat N|n)\bigr) +
E_n\bigl[\var(\hat N|n)\bigr].
\end{equation}
The first term on the right-hand side can be estimated
by $\sum_{i=1}^n (1-w_i)/w_i^2$, where
$w_i=1-e^{-\hat\rho_{i}\hat\gamma_{i}\hat\Omega(\tau)}$.
The second term can be computed approximately by means of the
delta method, where we have
%
%e4.22 #&#
%
\begin{equation}
\label{seN} \widehat{E_n\bigl[\var(\hat N|n)\bigr]} \to
\bigtriangledown g(\hat\theta)' J(\hat\theta)^{-1}
\bigtriangledown g(\hat\theta)
\end{equation}
with $\hat N=g(\hat\theta)$, $\hat
\theta$ being the maximum likelihood estimate of the parameter vector
and $J(\hat\theta)$ its information matrix. An expression for
$\bigtriangledown g(\hat\theta)$ is cumbersome, but the latter can be
easily derived using numerical differentiation methods. For a similar
strategy refer also to \citet{bohnvand09}.
%A detailed derivation of the latter for Gamma distributed random
%effects is given in Appendix \ref{seN}.

%s5 #&#
\section{Data description and analysis}
\label{data}

In Italy use or possession of a small amount of drugs may be
punished by administrative sanctions. If it is the first or
second offense, the Prefect may only issue a warning (Art. 75 modified
by law
$49/2006$). According to the Central Statistics Office of the Interior Ministry
(DCDS), about $30\mbox{,}000$ interviews are conducted in the presence of
Italian Prefects each year [\citet{EMCDDA2009}].
Following these interviews, around $25\mbox{,}000$
individuals (aged from 10 to 64 years) are formally warned to refrain
from further use of narcotic substances.
During our observation period, starting in 2006 immediately after
enforcement of
the new law, the DCDS
collected a database with information on subjects reported for
breaking the drug law (Art. 75 on personal use). Individuals are
identified with their
fiscal code (FC), which is the Italian equivalent of a social
security number in the USA. The database includes a record for each capture,
with date and time of detected abuse, and some information
about the subject including the FC for identification.
We restrict to the adult population, that is, subjects aged 18 or more.
In this work we use information related to the
gender, district of residence (South, Central,
North--West or North--East Italy, where subjects captured on the islands
separated from the Italian mainland are included as customary in the
Southern district), age at the start of the observation period and
its square.
%In our models we will also include the square of the age,
%to take into account a possibly nonlinear effects of age
%on the log-risk of capture.

A preliminary issue regards closure of the population of interest,
which is achieved here by restricting the observation period to 2 full
years. Given that the duration of
cannabis use is often several years [\citet{APD2009}], and that
frequent users who are most at risk of being captured are also
less likely to cessate
[\citet{chen1998predictors}], our study period may be a reasonable choice.
We have also performed a simple sensitivity analysis, by repeatedly
estimating the population size after truncation of the observation
period at shorter time horizons (6, 12 and 18 months).
All resulting estimates are rather close to those reported below.

%
%t2 #&#
\begin{table}
\tablewidth=244pt
\caption{Marginal and conditional counting
distributions for cannabis~users}% reported under Art. 75
% modified by Law 49/2006.}
\label{descript}
\begin{tabular*}{\tablewidth}{@{\extracolsep{\fill}}l c r r r r@{}} % centered columns (4 columns)
\hline
& & \multicolumn{4}{c@{}}{\textbf{Count}} \\ % inserts table heading
\hline Covariate & Description & 1 & 2 & 3 & 4 \\ % inserts table
%heading
[2pt]
Substance & Cannabis & 50,785 & 1124 & 60 & 4 \\
[2pt]
Gender & Male & 47,181 & 1035 & 55 &4 \\ % inserting body of the table
& Female & 3604 & 89 & 5 & 0 \\
[2pt]
District & South & 12,684 & 239 & 17 & 1\\
& Center & 18,427 & 411 & 22 & 0\\
& North--East & 7850 & 193 & 6 & 0\\
& North--West & 11,824 & 281 & 15 & 3\\
\hline
\end{tabular*}
\end{table}

Table~\ref{descript} shows the counting distributions over the categorical
covariates. The mean age is 25.01, with a standard deviation of
6.91.
In Table~\ref{c1c2} we report the log-likelihood and corresponding
population size estimate for general $M_{\mathrm{hotb}}$ models with
all predictors and different choices of $c_1$ and $c_2$.
On the basis of results in Table~\ref{c1c2}, we end up choosing $c_1=2$
and $c_2=\infty$.

%
%t3 #&#
\begin{table}%[b]
\tablewidth=244pt
\caption{Complete $M_{\mathrm{hotb}}$ models fit on the Italian cannabis users
data. We report the difference in
log-likelihood with respect to
the first model, which has log-likelihood $-453\mbox{,}196.3$}
\label{c1c2}
\begin{tabular*}{\tablewidth}{@{\extracolsep{\fill}}l c d{3.1} c@{}}
\hline%inserts double horizontal lines
$\bolds{c_1}$ & $\bolds{c_2}$ & \multicolumn{1}{c}{\textbf{log-lik}}
& \multicolumn{1}{c@{}}{$\bolds{\hat N/10^6}$} \\
\hline
1 & $\infty$ & 0 & 3.199 \\
2 & $\infty$ & 195.2 & 3.265 \\ % -453\mbox{,}001.1
3 & $\infty$ & 161.4 & 2.984 \\ %-453\mbox{,}034.9
4 & $\infty$ & 148.4 & 2.903 \\ %-453\mbox{,}047.9
1 & 2 & 33.9 & 3.079 \\
1 & 3 & 20.6 & 2.911 \\
2 & 3 & 171.6 & 3.150 \\
2 & 4 & 192.1 & 3.151 \\ %-453004.2
3 & 4 & 187.4 & 2.934 %-453008.9
\\
\hline
\end{tabular*}
\end{table}
%
%
%t4 #&#
\begin{table}[b]
\tablewidth=244pt
\caption{Comparison of $M_{\mathrm{hotb}}$ model with $c_1=2$,
$c_2=\Delta_b=\infty$ with nested models for the Italian cannabis
users data. We report the difference in
log-likelihood with respect to
the full model, which has~log-likelihood $-453\mbox{,}001.1$. All likelihood
ratio tests would lead~to~$p<0.0001$}
\label{comparesubmodels}
\begin{tabular*}{\tablewidth}{@{\extracolsep{\fill}}l d{3.1} r @{}}
\hline%inserts double horizontal lines
\textbf{Model} & \multicolumn{1}{c}{\textbf{log-lik}}
& \multicolumn{1}{c@{}}{$\bolds{\hat N/10^6}$} \\
\hline
$M_{\mathrm{hotb}}$ & 0 & 3.265 \\
$M_{\mathrm{otb}}$ & -13.8 & 2.417 \\ %submodels
$M_{\mathrm{htb}}$ & -28.4 & 3.254 \\
$M_{\mathrm{hob}}$ & -12.0 & 1.558 \\ %submodels
$M_{\mathrm{hot}}$ & -28.1 & 2.368  %submodels
\\
\hline
\end{tabular*}
\end{table}

In order to confirm the presence of all four sources
of heterogeneity with these data, we compare in Table~\ref{comparesubmodels} the chosen
$M_{\mathrm{hotb}}$ model to submodels. Each likelihood ratio test of the full
versus nested models has $p<0.0001$.
%On the basis of results in Table \ref{comparesubmodels}
%we confirm our choice of the full model $M_{hotb}$ with $c_1=2$.
There are a few features that we should notice:
first of all, the likelihood of model $M_{\mathrm{hot}}$ is very low when
compared to models including a behavioral effect. This allows us to
conclude that there is strong evidence in favor of the behavioral
effect. The peculiar behavioral effect detected can be explained as
a purely random variation, that is, a feature of the counting
distribution, or by considering that more serious legal consequences
are expected after the second time a subject is reported.
In the first case, we note from Table~\ref{descript} that there is
some sort of \textit{one-inflation} in the counting distribution, that
is, subjects are mostly captured only once. This may be a consequence of
the fact that the probability of repeatedly meeting the
same person by chance alone is low, at least in
large cities and in general within our short observation period.
In the second case, it can be noted that after the second capture
under the new Italian law, subjects may be submitted
to mandatory psychotherapy, they may be revoked their
driver's licence and entry visa when foreigners, and they may have to
pay a
fine and/or participate in treatment programs.
On the other hand, the first and second time a subject is identified
as a user have most likely the same
consequence of a warning by the judge. Identifications for breaking
other laws or other articles of the same law do not count toward
legal consequences arising at the third capture.
It is not surprising
then that subjects may be ``trap-shy'' after the second capture
(compare with estimate
$\hat\phi$ in Table~\ref{resMhotb} below). What we cannot say with
available information is if this is a feature of the counting
distribution which is due only to chance or if it is a feature of the
subjects being captured. In the second case, we also cannot say
if repression works, that is, subjects actually
quit using cannabis after the second time they are reported or if
they only start using it at home and have friends buy and carry it for
them.

%
%t5 #&#
\begin{table}
\tablewidth=240pt
\caption{Results of $M_{\mathrm{hotb}}$ model with $c_1=2$,
$c_2=\Delta_b=\infty$}
\label{resMhotb}
\begin{tabular*}{\tablewidth}{@{\extracolsep{\fill}}l d{3.3} d{5.3}@{}}
\hline%inserts double horizontal lines
\textbf{Parameter} & \multicolumn{1}{c}{\textbf{Estimate}} & \multicolumn{1}{c@{}}{\textbf{Std. Err.}} \\
\hline
$N$ & \multicolumn{1}{c}{3\mbox{,}265\mbox{,}071} & 91\mbox{,}247.39 \\
$\alpha$ & 129.69 & 0.56\\
$\phi$ & 0.20 & 0.01 \\
$\Omega(\tau)$ & 0.007 & 0.001 \\
$\beta(\mbox{Age})$ & -0.04 & 0.04 \\
$\beta(\mbox{Age}^2)$ & -0.03 & 0.05 \\
$\beta(\mbox{Female})$ & -0.65 & 0.11\\
$\beta(\mbox{North--East})$ & -2.13 & 0.08\\
$\beta(\mbox{North--West})$ & -2.20 & 0.07\\
$\beta(\mbox{South})$ & -2.03 & 0.07\\
\hline
\end{tabular*}
\end{table}

We also have evidence in favor of both observed and unobserved
heterogeneity, and also the assumption of time-homogeneity
shall be rejected.
Even if we obtain maximum likelihood estimates differently, most
of the models in Table~\ref{comparesubmodels}
correspond to a generalization of models in \citet{Chao2002}
(e.g., $M_{\mathrm{otb}}$ and submodels) and
\citet{Yip2007} (e.g., $M_{\mathrm{htb}}$ and submodels), allowing for a delayed onset
of the behavioral response.
For our selected model $M_{\mathrm{hotb}}$ with $c_1=2$ we
report in Table~\ref{resMhotb} the parameter estimates and standard errors.

Our final estimate for the population size of our catchable population
is slightly less than 3.3 million, which accounts
for approximately 8.9\% of Italians aged 18 to 64, and 7\% of the
entire adult population. In order to obtain an indirect estimate for
all cannabis users, we can use results in \citet{rossreyzuli11},
who claim that around two-thirds of the population of cannabis users
is catchable under Art. 75. Consequently, an indirect estimate for the
population of adults who have used cannabis at least once in Italy is
given by 5 million,
which accounts for 13.5\% of Italians aged 18 to 64,
or 10.6\% of the entire adult population.

Unobserved heterogeneity is relatively weak, even if present.
For what concerns observed heterogeneity, we find that
the
risk of capture decreases with age and its square. Females are at a
lower risk of
capture than men, and the subjects in the central district of
Italy are at the highest
risk of being reported. Note that we can interpret these findings
as a differential in prevalence only under the assumption
that time-heterogeneity is independent of predictors;
otherwise the difference in prevalence could be explained
by different ability of the officers in different Italian regions.
Under this assumption, as age grows use of cannabis in the street is
likely more
limited, females tend to use cannabis less than men, and the
largest number of users can be found in the central region.

These
findings are reasonable, as in the southern regions of Italy the
use of drugs
is more limited due to sociocultural differences as compared to the
central and northern parts. In the North--East and North--West we have
both a slightly smaller population at risk and possibly a larger prevalence
of cocaine rather than cannabis use.
This assumption could be easily verified/relaxed by
stratifying our Cox-type model to have a different baseline for each
category of an observed covariate combination.

%
%t6 #&#
\begin{table}
\caption{Estimates of population size for cannabis data for some
time-homogeneous population size~estimators}
\label{compareth}
\begin{tabular*}{\textwidth}{@{\extracolsep{\fill}}l cd{2.2}d{3.2}@{}}
\hline
\textbf{Estimator} & \multicolumn{1}{c}{$\bolds{\hat N/10^6}$} &
\multicolumn{1}{c}{$\bolds{\mathrm{Std}.\ \mathrm{Err.}/10^5}$}
& \multicolumn{1}{c@{}}{\textbf{log-lik}} \\
\hline
$M_0$ & 1.11 & 0.31 & -53.66 \\
[3pt]
Chao [\citet{Chao1987}] & 1.20 & 0.36 & -15.37 \\
%Zelterman \citet{Zelterman1988} & 1222861 & \\
[3pt]
$M_h$ Poisson2 [\citet{rivebail07}] & 1.85 & 1.06 & -19.54 \\
$M_h$ Darroch [\citet{Darroch1993}] & 3.70 & 4.43 & -16.16 \\
$M_h$ Gamma3.5 [\citet{rivebail07}] & 7.32 & 13.54 & -15.60 \\
[3pt]
$M_{\mathrm{ho}}$ [\citet{bohnvand09}] & 1.14 & 0.05 & \multicolumn{1}{c@{}}{$-$5660}\\
\hline
\end{tabular*}
\end{table}
%

% and would be related to
%the idea of collapsibility \citet{vandhetal12}.
%We finally comment on $\hat\phi$, which is significantly smaller than
%1, telling us that after the second capture the probability of being
%captured decreases strongly. As already noted,
%subjects become ``trap-shy'' after the second capture.
We conclude with a comparison with other estimators of the
population size, which can be found in Table~\ref{compareth}.
We use the independence model $M_0$, the
Chao [\citet{Chao1987}] %and Zelterman \citet{Zelterman1988}
lower bound estimator and other $M_h$ models, with the
assumptions for
the mixing distribution available in the function \texttt{closedp.0} in
the \texttt{R} package \texttt{Rcapture}.
%Note that the AIC values in Table \ref{compareth} are on
%a different scale with respect to those of our models in the previous
%tables, and should be used only within Table \ref{compareth}.
In Table~\ref{compareth} we obtain rather different estimates of the
population size, particularly relative to
the estimated standard errors. We
believe this happens when there are biases in population size
and standard error estimates. In this case,
any model not including the four sources of
heterogeneity may be misspecified for the data at hand.
Among the $M_h$ models, the largest likelihood arises
with the mixing distribution Gamma3.5.
The latter seems to grossly overestimate the population
size. It is not surprising that Gamma3.5 leads to overestimating the
population size, as with this mixing distribution
the capture probabilities are not bounded from below. See, for instance,
\citet{bailrive07} on this point and also for a detailed description
of the three mixing distributions used in Table~\ref{compareth}.
The $M_h$ Darroch model gives the closest estimate to our final
estimate of 3.3 million, but
with a standard error that is almost five times larger than ours.
Further, the latter would not be used in practice, as $M_h$
Gamma3.5 yields a larger likelihood with the same number of
parameters.
We also include the \citet{bohnvand09} model, an $M_{\mathrm{ho}}$
model generalizing that of \citet{Zelterman1988} and therefore
providing an efficient lower bound for the population size.
The resulting estimate is comparable to that obtained with the Chao lower
bound, with a smaller standard error due to the inclusion of
covariates.

Finally, we compare with two indirect estimates.
The first is given in \citet{rossi2012} using a dealer/consumer
ratio as proposed in \citet{bouctrem05}, and
estimates as 3.5 million the entire population of cannabis users.
The second is the official estimate of 5.5 million for
those who have used cannabis at least once in Italy, given by the
\citet{APD2009} and \citet{EMCDDA2010}.
As it is estimated that about 85\% of all
marijuana users are aged 18 or more [\citet{EMCDDA2010}],
it can be said that from \citet{rossi2012} we have an estimate of
about 3 million for the adult population
and that the official estimate is slightly under 4.7 million.
We believe these numbers to be
comparable with our indirect estimate of 5
million, and our results confirm the general
idea that official estimates may be slightly underestimating
the population size [e.g., \citet{rossi2012}].

%s6 #&#
\section{Discussion}
\label{concl}

We have proposed a general framework for continuous time
capture--recapture. Our model includes time-heterogeneity, unobserved
subject-specific heterogeneity, observed subject-specific
heterogeneity and behavioral response to capture. Classical behavioral
response has been generalized, allowing the user to specify a delayed
onset and a finite time memory.

In our application we
have found predictors that are able to explain some
heterogeneity, but we also have unobserved heterogeneity. A
model including both observed and unobserved
heterogeneity is therefore needed for these data. Unfortunately,
inferential approaches already available cannot be directly extended
to this case. Our EM-type estimation approach
can be easily adapted to any of the submodels of
$M_{\mathrm{hotb}}$.
Under the assumption of Gamma distributed random effects we
derived closed-form expressions for some of the quantities involved,
which greatly speeds up our algorithm.
We also found the
score in closed form and have made use of numerical differentiation
only to compute its first derivative.
The accuracy of
numerical first derivatives is much better than the accuracy of
numerical second derivatives, which we do not need in this paper.

We have developed a challenging application, with a small
sampling fraction and many sources of heterogeneity.
Available estimates of the number of cannabis users in Italy
are mostly based on chemical analyses of waste
waters or on consumer/dealers ratios. In our application we developed
a direct estimate of the population size of
adults ($\geq$18) who buy, carry and/or use cannabis in the streets
in Italy.
It shall be noted that our population includes only subjects
found in possession of less than 500 mg of cannabis. Possession of
more than 500 mg or for the purpose of
trafficking, selling, trafficking and cultivation are offences under
different articles of the Italian law, and offenders are not included
in our sample. Our final estimate for the population size is about 3.3
million, with a standard error of about 91 thousands.
Our final indirect estimate for the entire population of adult
cannabis users is 5 million.
It shall be
furthermore noted that subjects in possession of more than 500 mg,
trafficking, selling and/or cultivating are commonly estimated to be
around 200 thousand [\citet{rossi2012}].
This number is smaller than the width of our
estimated confidence interval, hence, even including these subjects, our
final estimates can be thought to be approximately the same.

With our age restriction we exclude an important
subpopulation, given that younger people are often the target of
prevention policies and delays in age of use may be associated
with lower risks [e.g., \citet{fergusson2006early}].
%The proposed model is valid
%only for approximately close populations, hence we must reduce as much
%as possible the possibility of new entrants.
Nevertheless, patterns of abuse, officer policies and prefect behavior
are different for underage users, hence, it is important that these two
populations are separately investigated. See also \citet
{kandel1984patterns} and
\citet{Ellickson2004} for a discussion on patterns of drug use.
Investigation of the under-18 population should in our opinion
be performed with open population models,
which are beyond the scope of this paper. Open population
models, furthermore, need a precise assessment of date of first
and very last use at least for some subjects, which may difficult to
measure without bias.

There are further limitations in our data set.
Given our catchable population is made
only of subjects who use, buy or
carry cannabis outside their apartment, there may be
concerns about the
interpretation of the estimated behavioral effect. We have
found that the risk of being captured decreases abruptly at the
second capture time. Even if this feature is not purely random,
it may be that after the second capture
most users do not actually quit, but start using drugs
at home instead of outside and have someone else buy for them.
We, furthermore, as often happens with capture--recapture
experiments, cannot guarantee that the population is closed.
A final limitation regards the limited information we have: we could
not take into account covariates related to the officers, and we do
not have information about time of day of each event. Therefore, we
cannot distinguish between different habits (e.g., day and night
users). We also do not have information regarding intensity of use and
arrests for other crimes, which directly affect the risk of capture.
%Our model takes anyway into account
%unobserved heterogeneity, which may arise due to these
%missing variables.

We conclude giving a brief account of possibilities for further work.
Our strategy for computation of standard errors may
yield estimates that are somewhat biased downward, given that we are
not taking
into account uncertainty brought about by model search. This is a
primary issue for further work.
Our model could also be extended to include
nonparametric baseline functions specific to known or even
unknown blocks of subjects. Finally, it could also be extended
to take into account spatial dependence.
%A different data
%collection scheme may also be used in the future, which
%will allow to define and fit open population models.

\section*{Acknowledgments}

The authors are grateful to five referees and the Editor (Professor T.
Gneiting) for very kind and detailed comments which helped improve the
presentation; and to Professor Carla Rossi for suggestions on an
earlier draft. This work has been developed within the framework of the
EU project JUST/2010/DPIP/AG/1410: ``New methodological tools for
policy and programme evaluation,'' with the financial support of the
Prevention and Information Programme of the European Commission. The
contents of this publication are the sole responsibility of the authors
and can in no way be taken to reflect the views of the European
Commission.

\begin{supplement}%[id=suppA]
\stitle{Derivation of (\ref{eqphi}), (\ref{eqO}) and (\ref{eqbeta}).}
\slink[doi]{10.1214/13-AOAS672SUPP} %[doi,text={...}] - jei reikia
%suskaldyti doi
\sdatatype{.pdf}
\sfilename{aoas672\_supp.pdf}
\sdescription{Derivatives of the log ECCL.}
\end{supplement}

% imsref loaded by lrinkeviciute, 2013-09-09 15:07:32
% imsref loaded by lrinkeviciute, 2013-09-09 15:19:13

\printaddresses

\end{document}